\RequirePackage{lineno}
\documentclass[aps,prl,preprint,superscriptaddress]{revtex4}

\usepackage{graphicx}
\usepackage{bm}
\usepackage{amsmath}
\usepackage{amssymb}
\usepackage{hyperref}
\usepackage{xcolor}

\begin{document}
\title{Lubricity of graphene on rough Au surfaces}
\author{Zhao Wang}
\email{zw@gxu.edu.cn}
\affiliation{Guangxi Key Laboratory for Relativistic Astrophysics, Department of Physics, Guangxi University, Nanning 530004, P. R. China.}

\begin{abstract}
This paper studies the lubricating properties of graphene on randomly rough Au surfaces in sliding nanofriction using molecular dynamics. It is shown that the friction and the consequent heat dissipation decrease more than an order of magnitude in the presence of graphene. The performance of graphene nanoribbons as lubricants is, however, limited because of detachment and displacement at the interface. Sliding contacts lubricated with a stretched graphene sheet exhibit low friction, but possibly also low structural stability. This suggests that the graphene-substrate adherence could be crucial for the lubricity of two-dimensional materials on rough metal surfaces.
\end{abstract}
\maketitle

\section{Introduction}
Friction is an unavoidable factor in mechanical systems. Metal-metal contacts, in particular, often exhibit strong adhesion and therefore high friction, leading to increased energy consumption, fast wear, and ultimately reduced lifetimes in mechanical components. Graphene, a monoatomic layer of carbon, is considered to hold great promise to reduce wear and friction by lubricating metal surfaces \cite{Berman2014}, thanks to its chemical inertness \cite{CastroNeto2009}, extreme mechanical strength \cite{Lee2010} and peculiar two-dimensional (2D) structure \cite{Wang2011}. For instance, Berman \textit{et al.} recently reported that graphene reduced wear and friction at the interface between steel surfaces by about $4$ orders of magnitude and by a factor of $6$, respectively \cite{Berman2013}. In particular, a single layer of graphene was found to resist loads for thousands of friction cycles in engineering-scale experiments \cite{Berman2014a}. Understanding the correlations between graphene's microscopic structural features and its lubricating behavior is therefore of high interest for the development of ultra-low-friction systems based on 2D materials.

Recently, Egberts \textit{et al.} performed nanofriction experiments by sliding a Si atomic-force-microscopy (AFM) tip over a graphene-covered Cu surface, and reported that the friction force can be reduced by a factor ranging from $1.5$ to $7.0$ \cite{Egberts2014}. Klemenz \textit{et al.} performed experiments and atomistic simulations of diamond-coated tips scratching over a graphene-covered Pt surface, and showed that friction remains low until the rupture of graphene \cite{Klemenz2014}. Smolyanitsky studied the temperature effect on the frictional properties of suspended graphene and demonstrated significant influence of thermal rippling \cite{Smolyanitsky2015, Smolyanitsky2014}. Li et al. showed that the time-dependent contact quality is critical for the friction on rough surfaces \cite{Li2016}. Although these results are revealing, the AFM-tip setup could be a limiting factor for studying lubrication of metal surfaces, since the strong interfacial adhesion (and associated wear) of metal-metal contacts cannot be correctly mimicked with AFM tips \cite{Szlufarska2008}. Similarly, most earlier theoretical works have resorted to atomistically flat surfaces that cannot be considered good models of realistic experimental conditions.

It is thus important to develop new microscopic test systems to address these issues and bridge the gap between microscopic investigations and macroscopic experiments. Here a novel set of quasi-three-dimensional molecular dynamics (MD) simulations is carried out to  study friction between graphene-lubricated Au surfaces with random roughness in the nanoscale. Au is chosen for testing because of its chemical inertness, easy experimental synthesis \cite{Nie2012} and superlubricity with graphene \cite{Kawai2016,Gigli2017,Cahangirov2013,Ouyang2016}.

\section{Methods}
\begin{figure}[htp]
\centerline{\includegraphics[width=11cm]{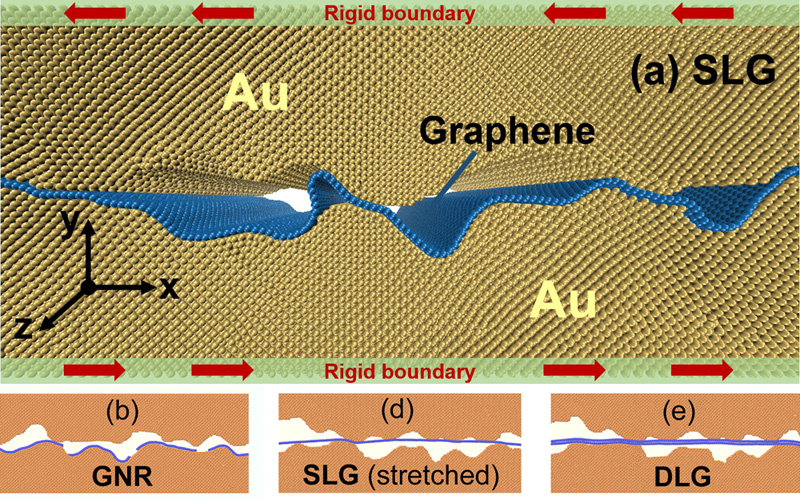}}
\caption{\label{F1}
Schematics of sliding nanofriction between Au films. The lower surface is coated with (a) a single-layered graphene sheet (SLG), (b) graphene nanoribbons (GNR), (c) a stretched SLG, (d) a stretched double-layered graphene sheet (DLG). The arrows show the sliding direction.
}
\end{figure}

We start by compressing two Au films against each other. Both of them are infinite (or more specifically periodic) in the plane normal to the compression direction, with a period of about $42.6\;\mathrm{nm}$. The films exhibit two-dimensional random surface profiles that are generated by a method and based on a discrete Fast Fourier Transform (FFT) algorithm \cite{Garcia1984}, in which the surface roughness is controlled by a correlation length and a root-mean-square height. Examples of the surface profiles are depicted in the Supplementary Information Fig. S1. After the two films are brought into contact in a given compression depth, the two surfaces slide along the $x$ axis in opposite directions. The sliding is realized by a controlled displacement of the rigid boundaries as shown in Fig.~\ref{F1}(a), with a constant relative sliding speed of $0.03\;\mathrm{nm/ps}$. Five series of essays are performed for a number of pairs of surfaces under different conditions listed in Table \ref{table1}. We note that it is a highly idealized situation that rough metal surface is fully covered by a continuous graphene sheet, while the case for wear of graphene is simulated with fragment graphene nanoribbons (GNR).

\begin{table}[h]
\caption{\label{table1} Samples tested under different conditions.}
\begin{center}
\begin{tabular}{ccc}
\hline \hline
sample & number of pairs & lubricant\\
\hline 
bare & 72 & clean \\
SLG  & 72 & graphene single-layer, Fig.\ref{F1}(a)\\
GNR  & 56 & graphene nanoribbons, Fig.\ref{F1}(b)\\
SLG (stretched) & 72 & stretched single-layer, Fig.\ref{F1}(c)\\
DLG  & 36 & double-layered, Fig.\ref{F1}(d)\\
\hline \hline
\end{tabular}
\end{center}
\end{table}

The parallel molecular dynamics package LAMMPS is used \cite{Plimpton95}. The interaction between the Au atoms is described by an embedded atom method (EAM) potential with the parameterization provided in Ref.~\cite{Grochola2005}, while that between the C and Au atoms is described by a Lennard-Jones potential with the parameterization proposed by Lewis \textit{et al.} \cite{Lewis2000} that enabled the prediction of 'ballistic' friction of Au nanoparticles on graphite \cite{Guerra2010}, which was later confirmed by experiments \cite{Lodge2016}. Recently, the combination of these potential functions was successfully applied to reveal graphene's superlubricity on a gold substrate \cite{Kawai2016}. Regarding the C-C atomistic interactions, The adaptive interatomic reactive empirical bond-order (AIREBO) potential \cite{Stuart2000a} is used instead of the Tersoff potential \cite{Tersoff1988a} that had been used previously \cite{Kawai2016,Guerra2010}. The reason is that the system undergoes severe plastic deformation when sliding: the AIREBO potential enables a smooth transition from long-range interaction to chemical bonding and thus affords a better description of possible bond formation and breaking \cite{Wang2009d,Wang2009carbon,Wang2011a,Wang2007,Wang2007a}. The equations of motion are integrated using the Verlet algorithm with a time step of $10^{-6}\;\mathrm{ns}$. A layer of heat-sink atoms are placed beneath each rigid boundary to maintain the system's temperature $T$ at approximately $300\;\mathrm{K}$ using the Nos\'{e}-Hoover thermostat. Note that the  Nos\'{e}-Hoover thermostat is a piston-type thermostat that makes temperature oscillate, and therefore is not suitable for being applied to all atoms at the contact. The atomic forces on the rigid boundaries are time-averaged and considered in relation to the sliding distance $\delta$ in order to compute the friction force $F_{x}$ and normal load $F_{y}$.

\section{Results and Discussion}

\begin{figure}[htp]
\centerline{\includegraphics[width=9cm]{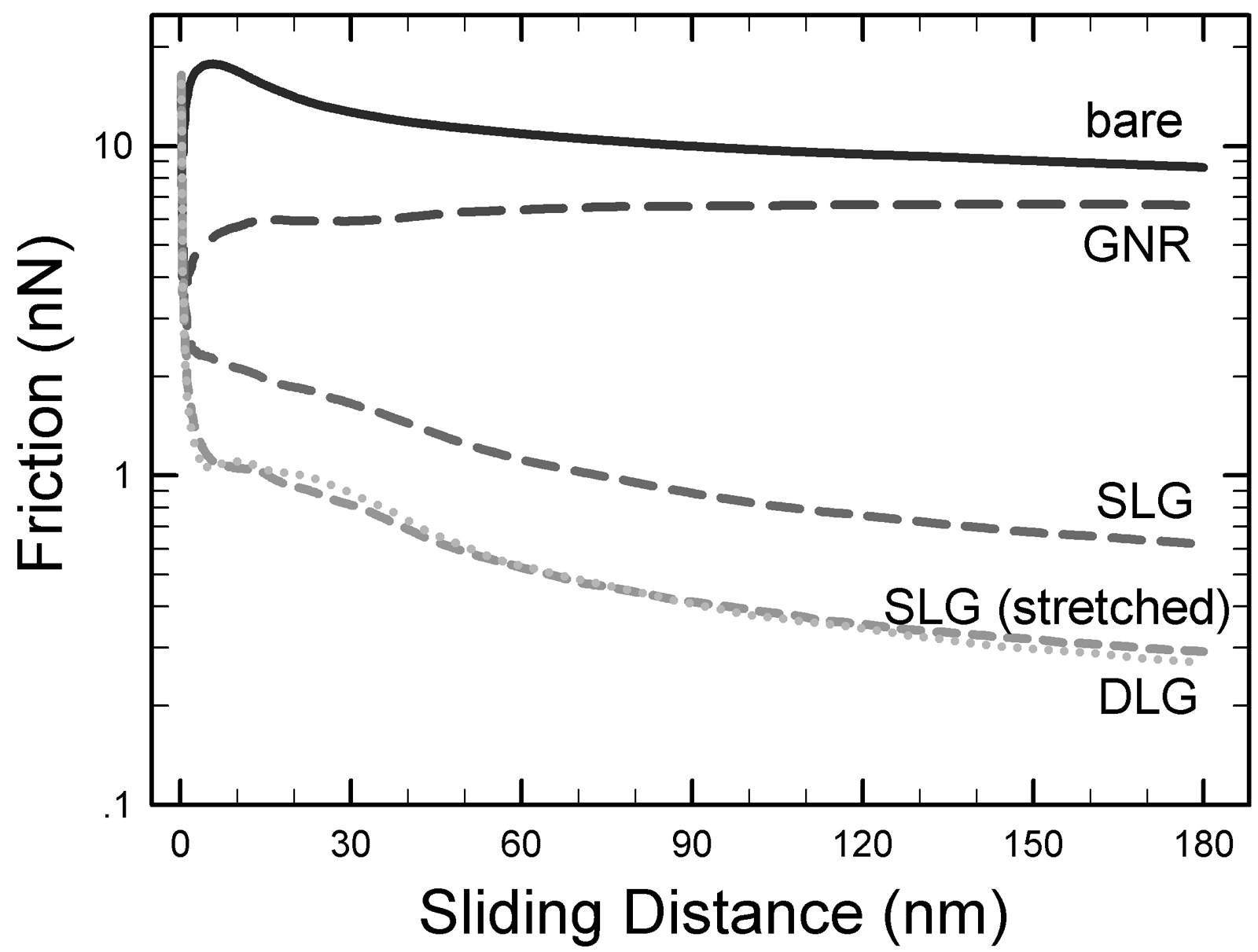}}
\caption{\label{F2}
Frictional force $F_{x}$ as a function of the sliding distance $\delta$. Line styles correspond to different cases tested.}
\end{figure}

Fig.~\ref{F2} shows the evolution of friction during sliding. It reveals that friction is high at the initial stages of sliding due to asperity deformation. With increasing sliding distance, the friction force decreases and tends to be constant when the surface is smoothed out by plastic deformation. The measured friction force scales from $0.3$ to $20.0\;\mathrm{nN}$ for a nominal surface of $36.3\;\mathrm{nm^2}$, comparable to experimental results that range from $1.0$ to $50.0\;\mathrm{nN}$ \cite{Egberts2014}. It is also apparent that friction decreases approximately by an order of magnitude in the case of full coating with a graphene sheet (SLG curve), but only by about $20\%$ in the case of GNRs. These contrasting lubricating behaviors are rather striking since the total initial surface coverages are similar in both cases. 

\begin{figure}[htp]
\centerline{\includegraphics[width=13cm]{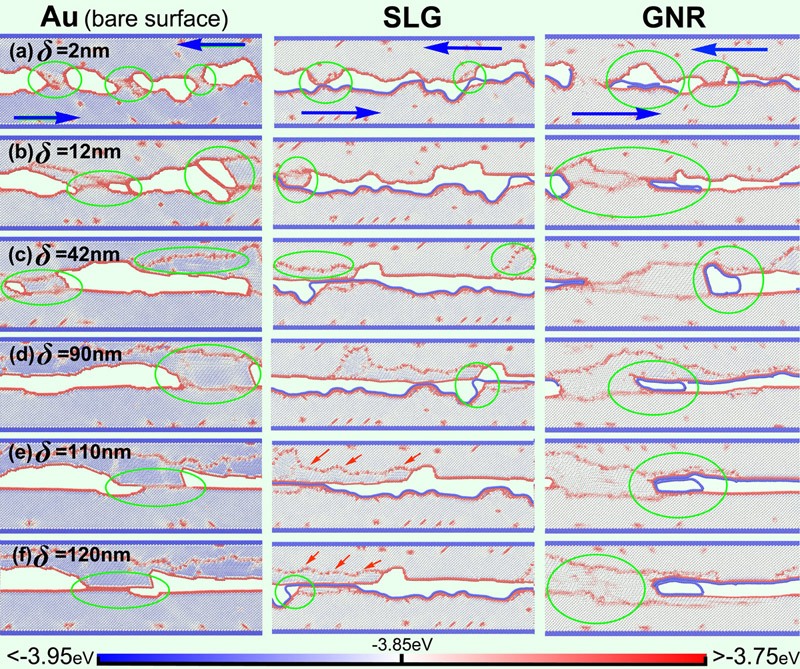}}
\caption{\label{F3}
Surface morphology evolution during sliding. The color scale denotes the atomic potential energy. Each column represents a different case. Nanofriction the stretched-SLG and DLG samples is shown in the Supplementary Information Figure S2.}
\end{figure}

The fragmented structure of GNRs is responsible for the deficient lubricity. Detachment and displacement of GNRs at the interface are observed when sliding occurs; therefore a part of the Au surface remains unprotected from direct Au-Au adhesion as shown in the \textit{right} panel of Fig.~\ref{F3}. In contrast, wear is significantly diminished in presence of a complete graphene sheet as shown in the \textit{central} panels of Fig.~\ref{F3}, although the Au films still undergo plastic deformation until the interface is polished. Given the fact that graphene has been proven to be impermeable to small molecules \cite{Rafiee2012,Bunch2008a}, the detachment and displacement of GNRs could be limited when the interface is filled with liquid or gas molecules. This provides an alternative response to the long-standing question of why graphene's lubrication performance is improved in liquid solutions \cite{Marchetto2017,Berman2013} but remains poor in vacuum \cite{Savage1948}, in addition to the superlubricity mechanism suggested by de Wijn \textit{et al.} \cite{Wijn2011}. Note that the detachment and displacement processes of GNRs can be more clearly seen in the simulation videos provided as part of the the Supplementary Information. 

Moreover, in Fig.~\ref{F3} there is strong metal-metal adhesion at the bare Au interface (\textit{left} panel) with shear strain evidenced by the formation of atom chains (b) and shear bands (b-d). Detachment and adsorption of graphene from one surface to another can be seen in the, \textit{central} panel of Fig.~\ref{F3} (highlighted by green circles). Defect nucleation zones beneath the interface are observed satisfying the criterion of incipient plasticity \cite{li02}.

\begin{figure}[htp]
\centerline{\includegraphics[width=9cm]{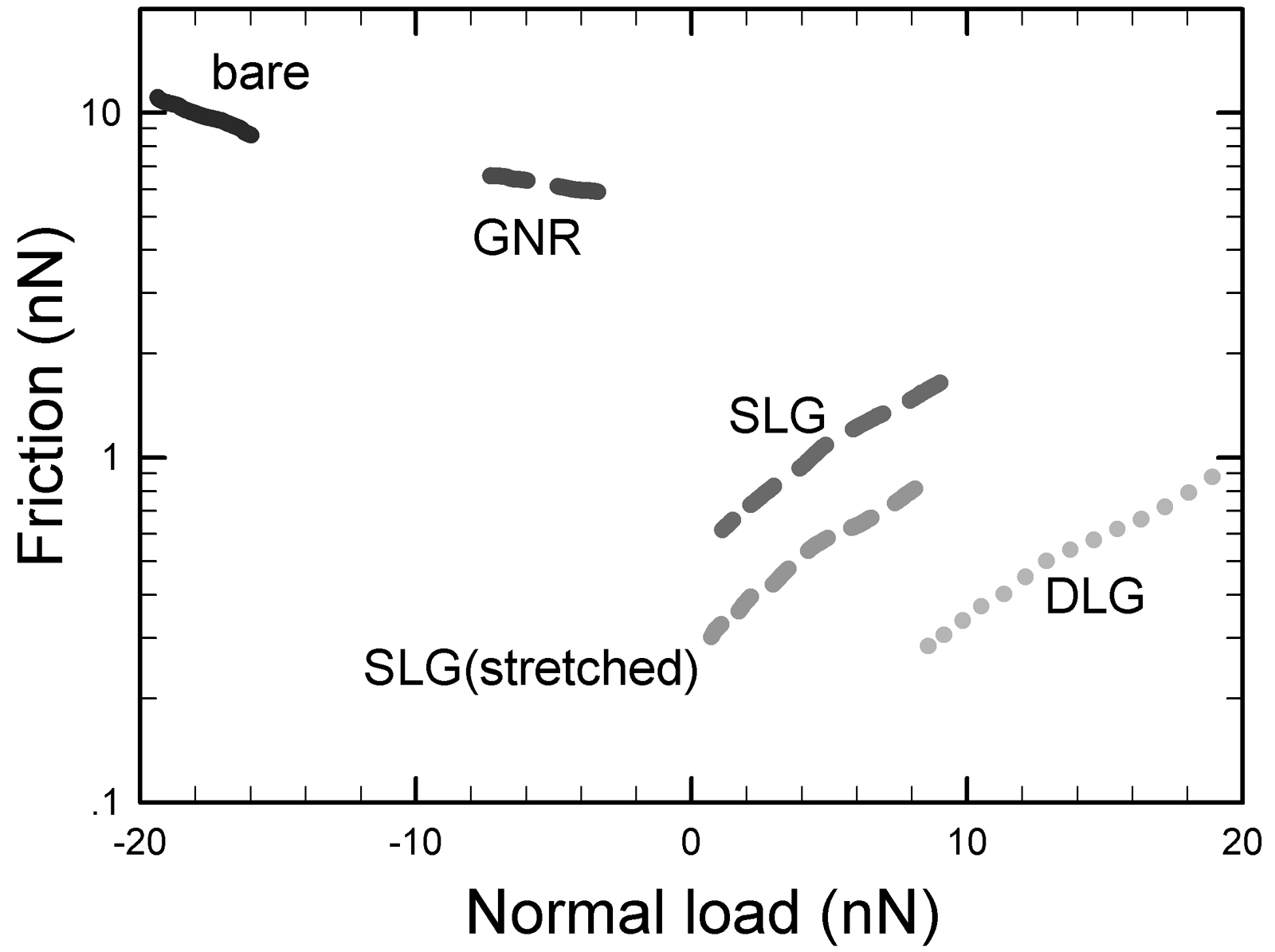}}
\caption{\label{F4}
Friction force $F_{x}$ as a function of normal load $F_{y}$.}
\end{figure}

It would be straightforward to discuss the lubrication performance in terms of the friction coefficient. However, since the contact between bare surfaces is strongly adhesive, the conventional Amonton's law remains barely applicable because of the negative normal load due to shear in a displacement-control scheme. This results in so-called ``negative coefficients of friction'' as observed in experiments on chemically modified graphite \cite{Deng2012} and other materials \cite{Thormann2013}. A phenomenological model was proposed to include the effect of adhesive friction \cite{Gao2004}, in which the friction force is a collection of both load- and adhesion-dependent contributions,  

\begin{equation} 
\label{Eq1}
F_{x}=\mu F_{y}+\sigma A,
\end{equation}

\noindent where $\sigma$ is the shear stress, $A$ stands for the contact area and $\mu$ is the coefficient of friction. For the bare and GNR-coated samples, the normal load is mostly measured to be negative as shown in Fig.~\ref{F4}, since the friction is dominated by the term $\sigma A$. Meanwhile, the true contact area cannot be precisely measured when the surfaces adhere together as shown in Fig.~\ref{F3}. This would make the contributions from individual terms in Eq.\ref{Eq1} undetermined. However, one can measure the slope of $F_{x}$ \textit{versus} $F_{y}$ curves in Fig.~\ref{F4} for large $\delta$ where $\sigma A$ tends to be constant. The values of $\mu$ are then estimated to be $0.105$, $0.0859$ and $0.0409$ for the SLG, stretched-SLG and DLG cases, respectively. With respect to an experimentally-measured $\mu=0.5$ for bare Au-Au contacts \cite{Gabel2001}, we estimate that the coefficient of friction decreases by a factor of about $4.8$, $5.8$ and $12.2$ for the SLG, stretched SLG and DLG samples, respectively. This is in qualitative agreement with Ye \textit{et al. }who have found by experiments and simulations that the friction decreases with increasing number of graphene layers with amplitude depending on the roughness \cite{Ye2017}.

\begin{figure}[htp]
\centerline{\includegraphics[width=9cm]{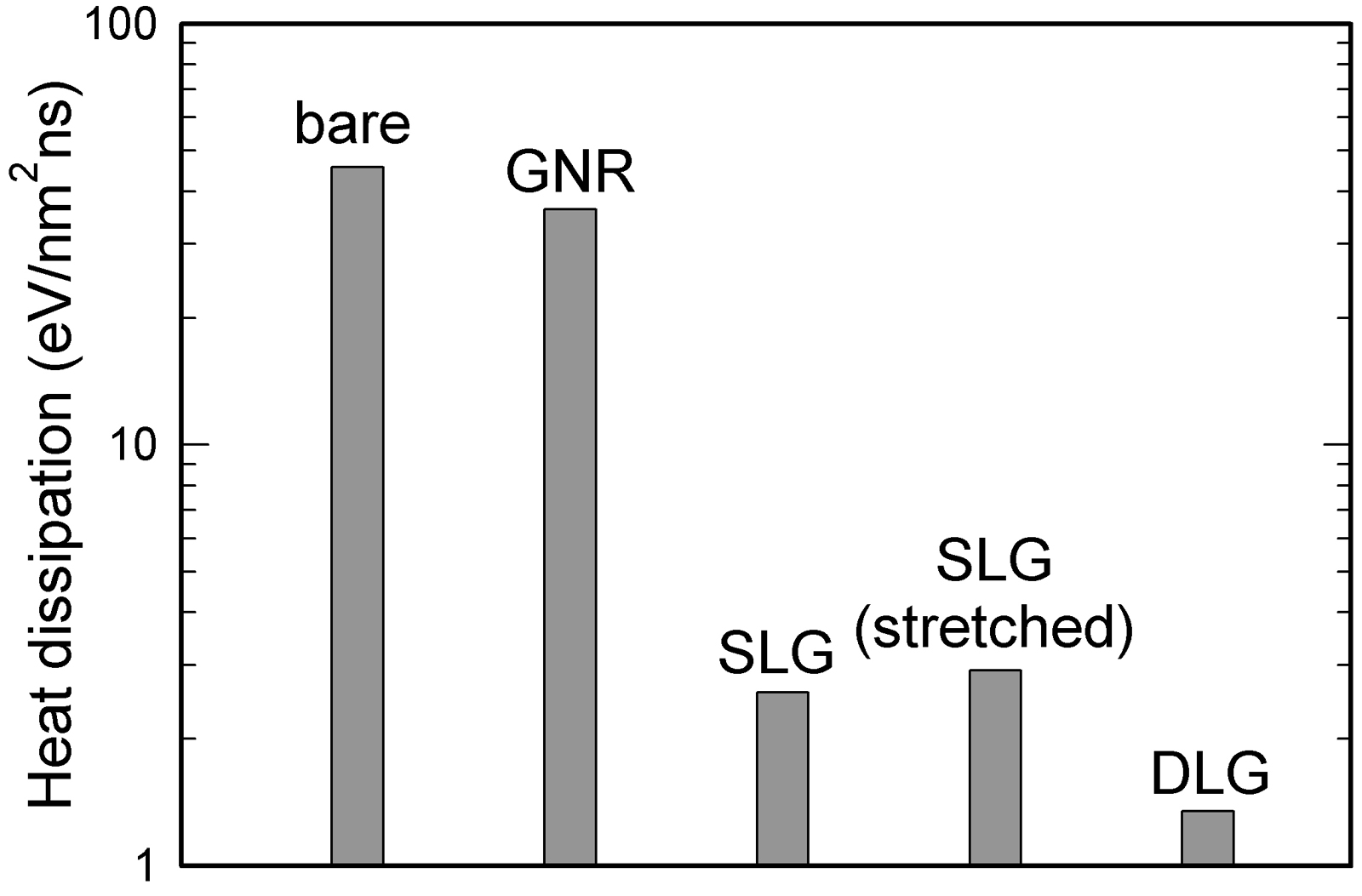}}
\caption{\label{F5}
Energy dissipated as heat per unit of time and cross section area.}
\end{figure}

Fig.~\ref{F5} shows the energy dissipation during sliding for different samples, which is computed by measuring the thermal energy transported through the heat sinks. Compared to the bare-surface samples, the energy dissipation is reduced by a factor of about $18$, $16$ and $34$ for the SLG, stretched-SLG and DLG samples, respectively. Although the friction force is found to decrease more with the stretched graphene sheet than the unstretched one (Fig.~\ref{F2}), the energy dissipation is however found to be slightly higher in the stretched samples. Reduced friction in the DLG sample may be due to the fact that the friction is between two graphene layers for the DLG case due to strong adhesion between Au and graphene, it is also probably due to that the deformation is compensated by local thermal fluctuation.

\begin{figure}[htp]
\centerline{\includegraphics[width=9cm]{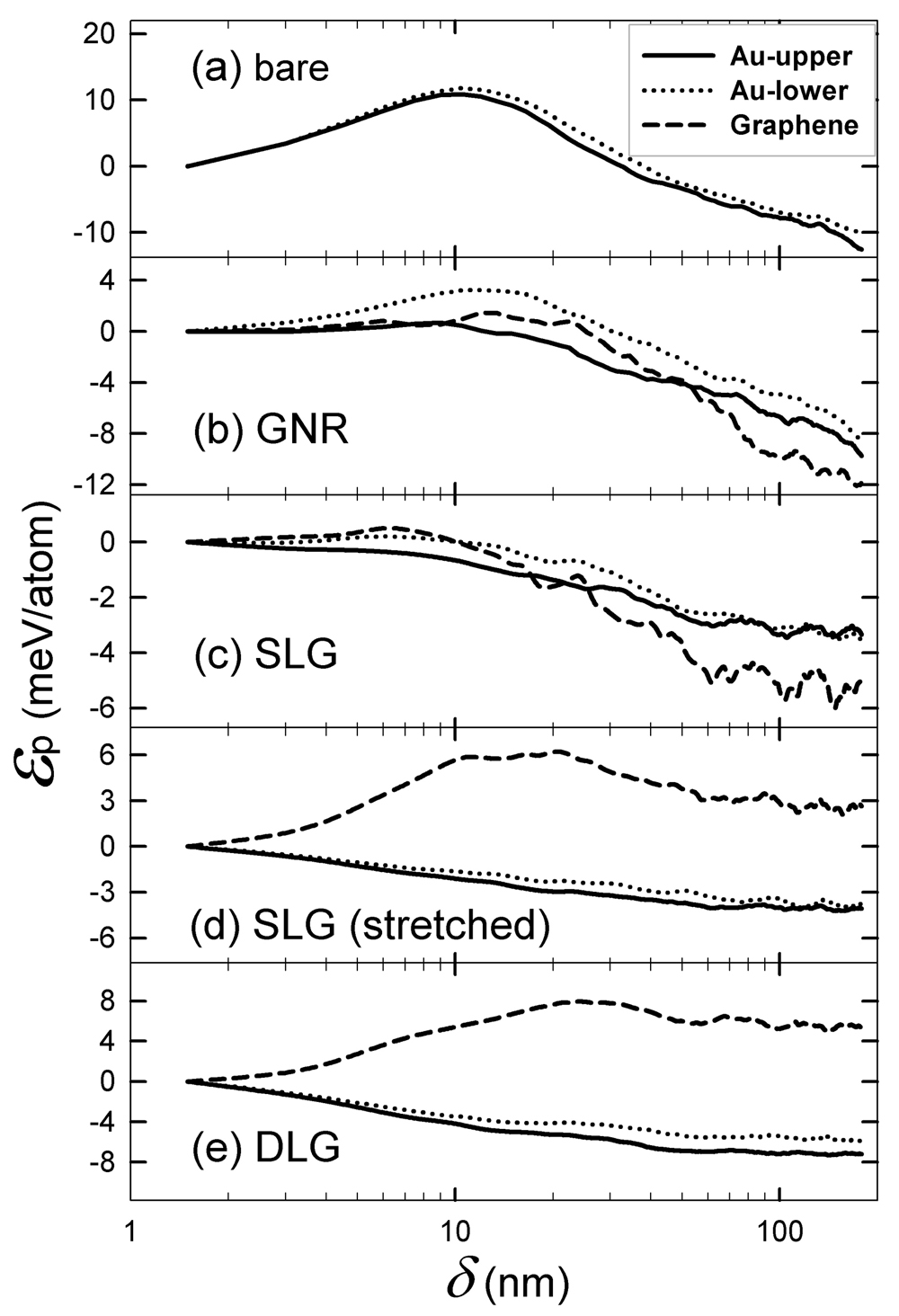}}
\caption{\label{F6}
Atomic potential energy as a function of sliding distance in logarithmic scale. The three line styles stand for the energy of the graphene,  upper and lower Au surfaces, respectively.}
\end{figure}

This is associated with the variation of potential energy of the system. Fig.~\ref{F6} shows the potential energy change during sliding. High potential energy peaks are observed at the initial stage of sliding in the bare and stretched-graphene samples. In such a case most deformation energy is ``absorbed'' by the graphene in the ``static'' regime of friction. Since the potential energy is an important index of structural stability, a lower lifetime can therefore be expected for the samples with stretched graphene, which exhibits low adherence to the substrate. Given that lubricants often work under cyclic loads in engineering applications, the adherence between the graphene and rough substrate is thus proven to be essential for maintaining the structural stability of the lubrication system.

\section{Conclusions}
Graphene exhibits remarkable lubrication performance on rough Au surfaces. The heat dissipation is lowered by more than an order of magnitude and the wear is significantly reduced by graphene lubricants. The friction coefficient is found to decrease from in the presence of SLG, stretched SLG and DLG, and the effect increases in this order. Energetic analyses suggest that higher structural stability can be expected from better graphene-substrate adherence. In contrast, detachment and displacement of graphene nanoribbons can be observed at the interface, causing the lubrication performance to be largely hindered when those are substituted for continuous graphene.

\section*{Acknowledgments}
Ju Li at MIT and Jesus Carrete at TU Wien are acknowledged for helpful discussions. This work is supported by the scientific research foundation of Guangxi University (XTZ160532) and Guangxi key laboratory foundation (15-140-54).

\end{document}